\documentclass[12pt]{iopart}

\usepackage{graphicx}
\begin{document}

%\preprint{APS/123-QED}

\title{Inverse Smith-Purcell effect near rough surfaces}

\author{  Zh.S. Gevorkian$^{1,2,*}$ and V. Gasparian$^{3}$}
\address{$^{1}$ Yerevan Physics Institute, Alikhanian Brothers St. 2,0036 Yerevan, Armenia. \\
$^{2}$ Institute of Radiophysics and Electronics, Ashtarak-2, 0203, Armenia.\\
$^{3}$ California State University, Bakersfield, USA\\
$^{*}$ gevork@yerphi.am}
\ead{vgasparyan@csub.edu}

%\author{Ann  Author}
 %\altaffiliation[Also at ]{Physics Department, XYZ University.}
%\author{Second Author}
%\affiliation{
%This line break forced with \textbackslash\textbackslash }

%\affiliation{ Second institution and/or address\\ }

%\pacs{87.14.gk}{DNA} \pacs{87.15.-v}{Biomolecules:structure and
%physical properties} \pacs{87.15.Zg}{Phase transitions}

\date{\today}

\begin{abstract}
Absorption of a photon by an electron moving parallel to a rough surface is studied.  In the weak scattering regime we have
evaluated the absorption probability of absorption of a single photon of energy $\omega$. It is shown the absorption probability with diffusional contribution becomes large by a $l_{in}/l \gg 1$ factor compared to the analogous result with the single scattering contribution.  The maximum of probability takes place for the infrared wavelengths and strongly depends on the particle energy. We also discuss the case of two-dimensional periodical surface profile and indicate optimal conditions for maximal absorption probability. The results can be used  in electron energy gain spectroscopy and in laser-driven acceleration.
\end{abstract}

%\begin{document}

\maketitle
\section{Introduction}
It is well known that a charged particle moving in the vacuum can not emit or absorb photons due to the energy-momentum conservation laws. On the other hand, emission or absorption is possible when particle moves in a medium or close to an interface. Cherenkov, Transition, Smith-Purcell radiations  are the examples of above mentioned emission (see, for example, \cite{RAD98}). During the recent years the inverse counterparts of the mentioned radiations  have been  observed \cite{Ed81}-\cite{Si05}. Earlier the inverse Smith-Purcell effect for periodical metallic gratings was theoretically analyzed in the sub-millimeter wavelength region \cite{BFS88}. The interest to these effects is largely motivated by a possibility of laser driven acceleration of charged particles. Besides, the inverse Smith-Purcell effect can be used in electron energy gain spectroscopy \cite{Abaj08}.

In the present paper we investigate the inverse Smith-Purcell effect, namely the absorption of a photon by an electron which moves parallel to a rough surface. To the best of our knowledge, no such calculations have been previously reported. The main difficulties with rough surfaces arise because it is more difficult to perform analytical
calculations when we deal with an arbitrary shaped profile of the grading for understandable reasons: there is no general algorithm to calculate the radiating part in the reflected waves. At present, most numerical simulations are available as one of the effective tools to analyze and to observe a variety of physical quantities such as electromagnetic fields as functions of time and space, power outflow, radiated intensity as a function of the radiating angle, etc. (see, e.g., \cite{new} and references
therein). Therefore any study of the inverse Smith-Purcell radiation from rough surfaces should be quite important and analytical
results are highly desirable. The purpose of the present work goes in this direction,
in the sense that we provide analytical expression for the absorption probability of a photon by an electron moving parallel to a rough surface.  In the diffusion regime we were able to obtain
a closed analytic expression for the absorption
probability, taking into account the diffusion contribution. We show that the diffusion contribution is dominant compared to the single scattering probability.

It has been considered recently, by one of the authors (Gevorkian), the radiation from a charged particle moving parallel to rough surfaces \cite{Gev10,Gev11}. The  averaged radiation intensity for a quite general surface random profile was directly calculated and it is shown that the main contribution to the radiation intensity is determined by the multiple scattering of polaritons induced by a charge on the surface.  We will develop an approach, following closely Refs.\cite{Gev10,Gev11}, which allows us to investigate periodical as well as random surface profiles, different materials from sub-millimeters to optics. We indicate necessary conditions for absorption to take place.

The plan of the work is as follows. In Sec. II we
briefly formulate the problem and introduce the basic equation for the absorption probability
of a single photon of energy $\omega$. In Sec. III we carry out an
exhaustive description of our two-dimensional rough surface and the analytical approach used in absorption probability
calculations. In Sec. IV we calculate the absorption probabilities with single and multiple scattering contributions.
It was shown that the diffusion contribution to absorption probability is the dominant one.
In Sec. V we discuss the utilization of the inverse Smith-Purcell effect for particle acceleration.
Finally, we summarize our results in Sec. VI.

\section{ Initial Relations}
Suppose that a fast electron moves on the positive $x$ direction parallel to a rough surface $xy$ at the distance $Z$ from it. Simultaneously a laser field of frequency $\omega$  falls down on the surface, see Fig.1.
\begin{figure}
\includegraphics[width=8.4cm]{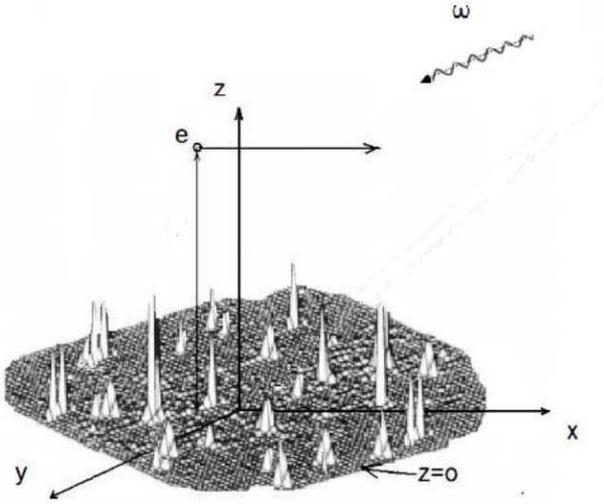}
\caption{Geometry of the problem. Electron moves parallel to the rough surface at the plane $xy$ which is illuminated by an external laser field.}
\label{fig.1}
\end{figure}

Electron wave function can be described as
\begin{equation}
\Phi(\vec r)=\frac{1}{\sqrt{L_x}}\varphi(Z,Y)e^{\frac{ip_ix}{\hbar}},
\label{wf}
\end{equation}
where $L_x$ is the system size in the $0x$ direction, $\varphi(Z,Y)$ is the wave function in the $zy$ plane, $p_i$ is the electron momentum along the direction of motion. After absorbing a photon electron momentum and energy become $p_f=p_i+\hbar q$ and
$E_f=E_i+\hbar qv$, respectively ($v$ is the velocity of the electron and a non-recoil approximation $\hbar q\ll p_i$ is assumed).
For a fast electron one can assume that the wave function $\varphi(Z,Y)$ remains unchanged during the interaction with the photon.  We will discuss conditions of applicability of this assumption below. Treating electron and photon as a quantum mechanical subjects, the absorption probability of absorption of a single photon of energy $\omega$ can be represented in the form \cite{Abaj08}
\begin{equation}
P(\omega)=\left(\frac{e}{\hbar\omega}\right)^2\left|\int dxE_x(x,Y,Z)e^{-i\frac{\omega x}{v}}\right|^2.
\label{abs}
\end{equation}
$E_x(x,Y,Z)$ is the electric field component along the electron motion direction that includes incident as well as scattered from surface fields and $Y,Z$ are the electron constant coordinates in the perpendicular to motion plane. If the incident field is a plane wave then it is easy to convince oneself that the incident part does not contribute to the integral. Therefore, for analytical evaluation of the integral, Eq.(\ref{abs}), the field $E_x$ will be substituted by the scattered one.  This will be done in the next sections, separately for the situations when  single and multiple scattering contributions are taken into account, while calculating the appropriate absorption probability.

\section{Scattered Field}
Dielectric constant of the system is described as $\varepsilon(\vec r)=\theta(z-h(x,y))+\varepsilon(\omega)\theta(h(x,y)-z)$, where $\theta$ is the step function and $\varepsilon(\omega)$ is the dielectric constant of the isotropic medium, $h(x,y)$ is the random profile of the surface. Assuming that $h$ is small and expanding  the $\varepsilon(\vec r)$ in powers of $h$ and keeping linear terms wee get $\varepsilon(\vec r)=\varepsilon_0(z)+\varepsilon_r(\vec r)$, where $\varepsilon_r(\vec r)=(\varepsilon-1)h(x,y)\delta(z)$. The function $\varepsilon_0(z)=1$ at $z>0$ and $\varepsilon_0(z)=\varepsilon(\omega)$ at $z<0$ describes the flat surface between vacuum and medium. Under this assumption, the scattered electric field can be represented as follows
\begin{equation}
E_{\mu s}(\vec r)=-\frac{\omega^2}{c^2}(\varepsilon-1)\int d\vec r^{\prime}G_{\mu\nu}(\vec r,\vec r^{\prime},\omega)h(\vec \rho^{\prime})\delta(z^{\prime})E_{\nu}^0(\vec r^{\prime},\omega).
\label{sc}
\end{equation}
$E_{\nu}^0(\vec r^{\prime},\omega)$ is the solution of Maxwell equation with the flat interface and because of the translational symmetry in the $xy$ plane it can be represented as follows: $E_{\nu}^0(\vec r)=e^{\vec k_{||} \vec \rho}E^0_{\nu}(z)$, where $\vec k_{||}$ and $\vec \rho$ are two dimensional vectors in the $xy$ plane. Green's function in Eq.(\ref{sc}) obeys the inhomogeneous Maxwell equation
\begin{equation}
\left[\varepsilon_0(z)\frac{\omega^2}{c^2}\delta_{\lambda\mu}-\frac{\partial^2}{\partial r_{\lambda}\partial r_{\mu}}+\delta_{\lambda\mu}\nabla^2+\varepsilon_r(\vec r)\frac{\omega^2}{c^2}\delta_{\lambda\mu}\right]G_{\mu\nu}(\vec r,\vec r^{\prime},\omega)=\delta_{\lambda\nu}\delta(\vec r-\vec r^{\prime}).
\label{gr}
\end{equation}
It is worth noticing that the presence  of the $\delta$-function in the expression of $\varepsilon_r$ will lead to the different values of any physical quantity at $z=0$, while evaluating the integral over $z$. To avoid the problem with discontinuous physical quantities at $z=0$ in our further calculations we will take their value at  $z=0^{+}$, see also, \cite{Mam75}. Such determination of integrals over $\delta$ functions give correct answers in the limit $|\varepsilon|\to\infty$.
Hence, substituting Eqs.(\ref{gr}) and (\ref{sc}) into Eqs.(\ref{abs}), one has
\begin{eqnarray}
P(\omega)=\left(\frac{e}{\hbar\omega}\right)^2(\varepsilon-1)^2\frac{\omega^4}{c^4}\int dxdx^{\prime}d\vec\rho_1 d\vec\rho_2
G_{x \nu}(x,Y,Z,\vec\rho_1,0^{+})G^{*}_{\mu x}(\vec\rho_2,0^{+},x^{\prime},Y,Z)\nonumber \\
h(\vec\rho_1)h(\vec\rho_2)
E^{0}_{\nu}(\vec\rho_1,0^{+})E^{0*}_{\mu}(\vec\rho_2,0^{+})
e^{-i\frac{\omega}{v}(x-x^{\prime})}.
\label{abs1}
\end{eqnarray}
 This is a general expression, independent of the model considered and can be applied for both, periodical and random grating cases. Below, we will analyze these cases separately (hereafter the sign $+$ is omitted).
%In further we  drop the sign $+$. We will consider both the periodical and random grating cases.
In the periodical grating case (photonic crystal) surface profile is a periodical function $h(\vec \rho)=\delta\cos\vec K\vec\rho$, where $\vec K=(2\pi/b,2\pi/d)$ is a two-dimensional vector and $b,d$ are the grating periods in the $x$ and $y$ directions, respectively. In the rough surface case $h(\vec\rho)$ is a Gaussian distributed random function. First let us consider the photonic crystal case. It is convenient first to present Green's function in the form
\begin{equation}
G_{\mu\nu}(\vec r, \vec r^{\prime})=\int G_{\mu\nu}(\vec p|z,z^{\prime})e^{i\vec p(\vec\rho-\vec\rho^{\prime})}\frac{d\vec p}{(2\pi)^2},
\label{fugr}
\end{equation}
where $G_{\mu\nu}(\vec p|z,z^{\prime})$ is the Fourier transform in the $xy$ plane. In the second step, let us assume
that the plane of incidence of external light is $xz$. Then the background electric field in Eq.(\ref{abs1}) that includes incident and reflected parts, takes the form: $E_{\nu}(\vec\rho,0+)=e^{ik_xx}E_{\nu}$, where $k_x=\omega\cos\theta/c$ and $\theta$ is the angle between the external photon momentum and electron velocity directions. Substituting expressions for Green's function, electric field and $h(\vec\rho)$ into Eq.(\ref{abs1}), one finds
\begin{equation}
P(\omega)=\frac{g_1\pi L_x}{2}G_{x\nu}\bigg(\frac{\omega}{v},\frac{2\pi}{d}|Z,0\bigg)G_{\mu x}^{*}\bigg(-\frac{\omega}{v},-\frac{2\pi}{d}|0,Z\bigg)\delta\bigg(k_x+\frac{2\pi}{b}-\frac{\omega}{v}\bigg)E_{\nu}E_{\mu}^{*},
\label{abs3}
\end{equation}
where $g_1=(e/\hbar\omega)^2\frac{\omega^4}{c^4}(\varepsilon-1)^2\delta^2$, $L_x$ is the system size in the $x$ direction and $\delta(k_x=0)$ was substituted by $L_x/2\pi$. In the weak scattering regime $(\varepsilon-1)^2\delta^2/\lambda^2\ll 1$ the Green's functions in Eq.(\ref{abs3}) can be substituted by the bare Green's functions. The latter quantities are the solutions of Eqs.(\ref{gr}) with $\varepsilon_r\equiv 0$ and were found in \cite{Mam75}. The existence of the $\delta$ function in Eq.(\ref{abs3}) sets the relation between  the external light wavelength, incident angle, electron velocity and the gratings period. Interestingly, they are related to each other in the same way as in the direct Smith-Purcell effect
\begin{equation}
\lambda=b\bigg(\frac{1}{\beta}-cos\theta\bigg),
\label{sp}
\end{equation}
with $\beta=v/c$. Note that the dispersion relation depends only on the grating period in the electron velocity direction. As it was is shown in \cite{Mam75} $G_{xy}=G_{yx}\equiv 0$ which for our problem means that a photon polarized in the perpendicular to incidence plane (s-polarization) can not be absorbed by the electron. Therefore we will consider the case when the incident photon is p-polarized. To simplify the problem consider the limit $|\varepsilon|\gg 1$. In this case the main contribution to the Eq.(\ref{abs3}) comes
from the term containing $G_{xz}$. The explicit expression of the bare Green's function is (see \cite{Mam75})
\begin{equation}
G_{xz}(\vec p|z,0)=-G_{zx}(\vec p|0,z)=-\frac{ip_x}{k^2}\frac{\varepsilon(\omega)qe^{iqz}}{k_1-\varepsilon(\omega)q},
\label{fugr2}
\end{equation}
where $q=\sqrt{k^2-p^2}$ if  $k^2>p^2$  and
$i\sqrt{p^2-k^2}$  if  $k^2<p^2$ and $k_1=-(\varepsilon(\omega)k^2-p^2)^{1/2}$. Substituting Eq.(\ref{fugr2}) into Eq.(\ref{abs3}), one comes after some algebraic manipulations to the following expression for $P(\omega)$

\begin{eqnarray}
&P(\omega)&=\frac{\pi g_1L_x}{2}\frac{c^2|\varepsilon(\omega)|^2(\gamma^{-2}+\frac{\lambda^2\beta^2}{d^2})|E_z|^2\delta(\frac{\omega}{v}-
\frac{2\pi}{b}-k_x)}{\beta^2\omega^2\left[(\varepsilon(\omega)\beta^2-1-\frac{\lambda^2\beta^2}{d^2})^{1/2}+i\varepsilon(\omega)\sqrt{\gamma^{-2}+
\frac{\lambda^2\beta^2}{d^2}}\right]}\times \\ \nonumber
&\times& \frac{\exp\left(-4\pi Z\sqrt{\frac{1}{d^2}+\frac{1}{\gamma^2\beta^2\lambda^2}}\right)}{\left[(\varepsilon^*(\omega)\beta^2-1-\frac{\lambda^2\beta^2}{d^2})^{1/2}-i\varepsilon^*(\omega)\sqrt{\gamma^{-2}+
\frac{\lambda^2\beta^2}{d^2}}\right]},
\label{complex}
\end{eqnarray}
with $\gamma=(1-\beta^2)^{-1/2}$. The amplitude of electric field includes both incident and reflected parts, $E_z=(1+r(\omega))E_z^1$, where $E_z^1$ is the amplitude of incident field and $r(\omega)$ is the reflection amplitude that goes to unity in the limit $|\varepsilon|\to\infty$. Because of the exponential function  absorption takes place for the electron distances from the surface satisfying the condition
\begin{equation}
4\pi\left(\frac{1}{d^2}+\frac{1}{\lambda^2\beta^2\gamma^2}\right)^{1/2}Z\ll 1
\label{dist}
\end{equation}
For the distances Eq.(\ref{dist}) electron transverse wave function $\varphi(Z,Y)$ remains unchanged during the interaction with the photon. We need this assumption for derivation of absorption probability Eq.(\ref{abs}).
 Essential absorption probability is achieved in the case when the imaginary part of $\varepsilon(\omega)$ is small compared to the real part. Such a situation occurs, for example, for noble metals {\it Au,Ag, Cu and etc} at the infrared wavelengths \cite{JC72}. As an example for gold at photon energy $\hbar\omega=1ev$, $Re\varepsilon=-70$ and $Im\varepsilon=6.27$. Assuming $\varepsilon(\omega)$ real, for the distances Eq.(\ref{dist}),one obtains
\begin{equation}
P(\omega)=\frac{\pi g_1L_x}{2}\frac{c^2\varepsilon^2(\omega)(\gamma^{-2}+\frac{\lambda^2\beta^2}{d^2})|E_z|^2\delta(\frac{\omega}{v}-
\frac{2\pi}{b}-k_x)}{\beta^2\omega^2\left[\varepsilon(\omega)\beta^2-1-\frac{\lambda^2\beta^2}{d^2}+\varepsilon^2(\omega)(\gamma^{-2}+
\frac{\lambda^2\beta^2}{d^2})\right]}.
\label{abs4}
\end{equation}
Since the absorption probability is positive one gets a condition on the particle velocity
\begin{equation}
\frac{v^2}{c^2}\leq \frac{(\varepsilon+1)}{\varepsilon-\frac{\lambda^2(\varepsilon+1)}{d^2}}
\label{vel}
\end{equation}
Maximum of $P(\omega)$ is achieved provided that the equality in Eq.(\ref{vel}) holds. It follows from Eq.(\ref{vel}) that the wavelength $\lambda\gg d$ is suppressed. Taking the limit $d \to \infty$ one returns,  as it should be expected, to the Smith-Purcell geometry with periodicity only in the direction of particle motion. In this geometry and for metallic gratings $(\varepsilon<-1)$ the quantity $c\sqrt{(\varepsilon+1)/\varepsilon}$ represents the velocity of induced  polariton on the surface. Hence, the condition Eq.(\ref{vel}), means that the maximum of absorption is reached if the particle velocity equals the polariton velocity.

\section{ Rough Surface}
The above consideration can be generalized to a case when a charged particle travels over a rough surface, by applying a recently developed approach to study the spectrum of radiation from a surface roughness \cite{Gev10,Gev11}. It can be shown that in this case absorption probability consists of two parts, each of them has different origin and must be evaluated separately. One is caused by the single scattering of polaritons another is caused by their diffusion, i.e., by the multiple scattering effect.  Using Eq.(\ref{abs1}), for the single scattering contribution to the absorption probability, one obtains
\begin{eqnarray}
P^{r}(\omega)=g_1\int dx dx^{\prime}d\vec\rho_1\vec\rho_2 G_{x\nu}(x,Y,Z,\vec\rho_1,0^{+})G_{\mu x}^{*}(\vec\rho_2,0^{+},x^{\prime},Y,Z)\\ \nonumber W(|\vec\rho_1-\vec\rho_2|)E_{\nu}(\vec\rho_1)E^{*}_{\mu}(\vec\rho_2)e^{-i\frac{\omega(x-x^{\prime})}{v}}
\label{single}
\end{eqnarray}
where $\delta^2W(|\vec\rho_1-\vec\rho_2|)=<h(\vec\rho_1)h(\vec\rho_2)>$ is the correlation function of the rough surface profile. We assume that the surface profile fluctuations are uncorrelated. This allows us to substitute the correlation function $W$ by $\delta$ function except the cases when finite correlation length is needed for divergence reasons, i.e. to avoid the divergences in integral calculations. Such an approximation is justified provided that $\lambda\gg \sigma$, where $\sigma$ is the correlation length of random surface profile fluctuations. Next, we evaluate  the integral Eq.(\ref{single}), by assuming $W(p)=\pi\sigma^2e^{-p^2\sigma^2/4}$. Then, taking the Fourier transforms, substituting Eq.(\ref{fugr2}) into Eq.(\ref{single}) in the limit $Z\to 0$ (more precisely $Z\ll \beta\lambda\gamma/2\pi$), we obtain the desired result for the absorption probability with single scattering contribution
\begin{equation}
P^r(\omega)=\frac{\pi L_xg(1+r)(1+r^*)|E_z|^2c}{\beta^4\omega}F(\varepsilon,\beta)
\label{single1}
\end{equation}
where
\begin{equation}
F(\varepsilon,\beta)=\frac{\varepsilon^2}{(\varepsilon^2-1)}\left[1+\frac{\omega\pi\sigma\varepsilon\beta}
{2c\sqrt{(1+\varepsilon)(1+\varepsilon\gamma^{-2})}}\right].
\label{endep}
\end{equation}
Note that in Eqs.(\ref{single1},\ref{endep}), like before, we assume that imaginary part of $\varepsilon$ is small and neglect it compared to real part. For the positive $\varepsilon$ (dielectrics), as it follows from $F(\varepsilon,\beta)$, the expression under the square root is always positive.
As for negative $\varepsilon$ (metals), it can be easily checked, that it leads to serious restriction on the energy of particle, i.e. $\gamma^2\leq -\varepsilon$. Maximum  absorption is achieved when the equality holds, i.e., $\gamma^2=-\varepsilon(\omega)$, see Fig.2.
\begin{figure}
\includegraphics[width=8.4cm]{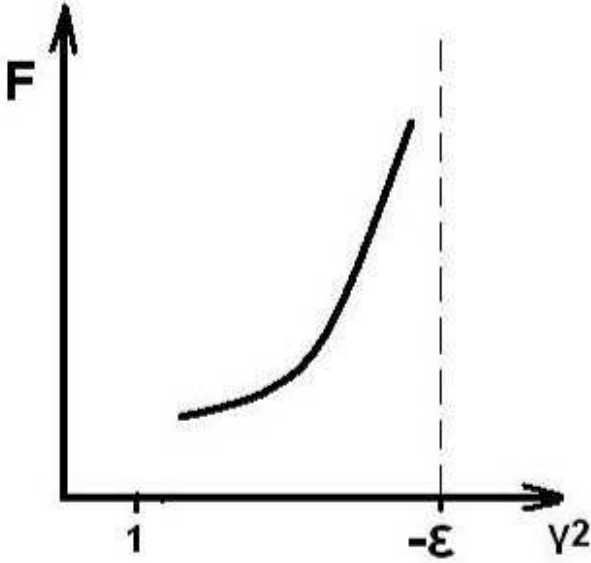}
\caption{Absorption probability dependence on the particle energy.}
\label{fig.2}
\end{figure}
For the negative $\varepsilon$ a plasmon-polariton is formed on the surface, see for example Ref. \cite{Ra88}. The pole at $p^2=\varepsilon k^2/(\varepsilon+1)$ in the Green's function Eq.(\ref{fugr2}) is  manifestation of the plasmon-polariton. It is scattered on the inhomogeneities and gives contribution to the scattered electric field in Eq.(\ref{abs}) and hence to the absorption probability $P(\omega)$. The expression (\ref{single1}) is  the plasmon-polariton single scattering contribution. To make a further analytical progress in the study of $P(\omega)$ we will assume that the following inequality is met: $\lambda<<l<<l_{in},L$, where $l,l_{in}$ are elastic and inelastic mean free paths of polariton on the surface. In other words we will assume that the condition of multiple or diffusional scattering of polariton are realized in the surface. In our calculations of the diffusional contribution to the absorption probability we follow closely Refs. \cite{Gev10,Gev11}. Further manipulations are completely analogous to those outlined in Refs.\cite{Gev10,Gev11} for the case of the radiation problem. Hence, here we present the final result without derivation by noting, that the diffusion contribution to absorption probability is the dominant one
\begin{equation}
P^{D}(\omega)=\frac{32}{3}\frac{l_{in}}{l}P^r(\omega).
\label{diff}
\end{equation}
Indeed, as seen from Eq.(\ref{diff}), the quantity $P^{D}(\omega)$ is proportional to $P^r(\omega)$, with prefactor $l_{in}/l$, which is the average number of polariton scatterings in the system. In the diffusion regime the ratio is large number, i.e $l_{in}/l\gg 1$, see also \cite{GG13}, justifying that the diffusion contribution is dominant. It is important to notice, that one of the advantages of the random surface profile is that the external light incident angle can be arbitrary instead the certain one in the periodical case.

For completeness , we also compare the absorption probability with the probability of emission of a photon by a charge particle moving under the same conditions, see Fig.1. The probability of emission of a photon of energy $\omega$ by an electron moving over a rough surface can be estimated as (following Refs.\cite{Gev10,Gev11})
\begin{equation}
P^e(\omega)\approx \frac{2e^2}{3\hbar c\beta^2}g_0(\omega)\frac{L_x}{Z}\frac{l_{in}(\omega)}{l(\omega)},
\label{emiss}
\end{equation}
where $g_0=(\varepsilon-1)^2k^4\delta^2\sigma^2$ and $Z\ll \lambda\beta\gamma/2\pi$ is the distance from the plane $z=0$.
Using Eqs.(\ref{single1}), (\ref{endep}), (\ref{diff}) and (\ref{emiss}) the ratio of probabilities can be estimated as
\begin{equation}
R=\frac{P^D(\omega)}{P^e(\omega)}\approx\frac{16Zc^3|E_z|^2(1+r)(1+r^*)}{\hbar \omega^4}F(\varepsilon,\beta).
\label{ratio}
\end{equation}
Now let us estimate numerically $R$. Before doing so, first we verify numerically the applicability of the diffusion approximation. Note, that in the weak scattering regime average mean free paths are described by the following expressions: $l=4|Re\varepsilon|/kg_0$ and $l_{in}=(Re\varepsilon)^2/kIm\varepsilon$ \cite{Gev10,Gev11}. For $Au$ at the photon energy $\hbar\omega=1eV$ $Re\varepsilon=-70$ and $Im\varepsilon=6.27$ , $r=r^{*}\sim 1$. Taking for the roughness parameters $\delta=10nm$ and $\sigma=100nm$ one gets $g_0\sim 3.13$ and $l\sim 14\lambda$ and $l_{in}\sim 124\lambda$. This means that the conditions $\lambda\ll l\ll l_{in}$ of diffusion of polaritons are realized in the system. Now taking electron energy $E=3.5Mev$ , $Z\sim\lambda\beta\gamma/2\pi$, laser power $|E_z|^2c\sim 10^{10}Wm^{-2}$ one finds from Eq.(\ref{ratio}) that $R\sim 18$.

Note that in contrary to the radiation case \cite{Gev11}, where maximum is achieved for short wavelengths (blue part of visible region), here maximum of probability takes place for infrared wavelengths. An interesting and difference feature of the absorption probability, compared to the radiation case is its  strong dependence  on the particle energy Fig.2. This dependence can be quantitatively measured and can be used for investigation of metal dielectric constant $\varepsilon(\omega)$ in the optical region.

 Concluding this section let us note that the ratio $R$ can be made essentially larger via increasing the laser power. This point is topical for the laser driven acceleration application of the inverse Smith-Purcell effect.
\section{Laser driven acceleration}
We now want to discuss  utilization  of the inverse Smith-Purcell effect for particle acceleration. Metal surfaces with rough or one-dimensional periodic gratings can not be used for acceleration purposes because of the restriction on the energy of particle (see Eq.(\ref{vel}). However, there is an important exception, when the strength of the electric field, that determines the absorption probability and were scattered from a metal surface, can be resonantly large. To illustrate this, we consider two-dimensional periodical grating case. The absorption probability, Eq.(\ref{abs4}), is straightforwardly applicable in this case.
 Rewriting the restriction condition on the energy, Eq.(\ref{vel}), in the form
\begin{equation}
\gamma^2\leq\frac{\frac{\lambda^2}{d^2}(\varepsilon+1)-\varepsilon}{1+\frac{\lambda^2}{d^2}(\varepsilon+1)},
\label{resen}
\end{equation}
it is easy to see, that the most favorable situation happens when the photon energy satisfies the resonant condition, i.e. the  denominator of Eq.(\ref{resen}) becomes zero
\begin{equation}
1+\left(\frac{2\pi c}{\omega d}\right)^2(\varepsilon(\omega)+1)=0
\label{res}
\end{equation}
Note that we have in mind optical frequencies for which $\varepsilon(\omega)$ is a large negative number. In this case the energy of accelerated particle can be very large. The absorption probability Eq.(\ref{abs4}) at the resonance photon energy will be large too.

The largeness of absorption probability is caused by the resonance enhancement of the scattered field  due to the surface plasmon-polaritons. Growth of probability is limited only by the losses in the optical region. In dielectrics restriction on energy is absent and they can be used in acceleration purposes, see for example, Ref.\cite{Co03}.
\section{Summary}
In conclusion, we have investigated the  absorption of a photon by an electron moving over a rough surface. Optimal conditions that include polarization of incident light, electron energy, material and grating types are indicated. In particular, it is shown that only p-polarized photon can be absorbed. For metallic surfaces and for relativistic particles two-dimensional periodical grating is preferable because of the restriction on the energy of particle. For dielectrics restriction on the energy is absent.

\subsection{Acknowledgments}
We are grateful to A.Akopian  and F.J.de Abajo for helpful comments. V.G. acknowledges partial support by FEDER and the Spanish DGI under project no.FSI2010-16430.
%\begin{thebibliography}{0}

% \Name{Author F., Author S. \and Author T.}
 % \REVIEW{Some Rev. A}{69}{1969}{9691}.

%\bibitem{b.b}
 % \Name{Author F. \and Author S.}
  %\Book{Some Book of Interest}
  %\Editor{A. Editor}
  %\Vol{9}
  %Publ{Publishing house, City}
  %\Year{1939}
  %\Page{666}.

%\bibitem{b.c}
 % \Editor{Editor A.}
 % \Book{Some Book of Interest}
  %\Vol{9}
  %\Publ{Publishing house, City}
  %\Year{1939}
  %\Section{A}.

%\end{thebibliography}

\end{document}